\documentclass{article}
\usepackage[latin1]{inputenc}

\makeatletter

\newcommand{\LyX}{L\kern-.1667em\lower.25em\hbox{Y}\kern-.125emX\@}

\makeatother

\begin{document}

\title
{The quadratic metric as only possible metric for }

\title
{the quantized space.}

\author{Roman G. Vorobyov}

\maketitle
\begin{abstract}
It is shown, that the space quantum existence ( SQE ) non-ambiguously determines
the metrical form for the space without a time, using weak condition of metrical
additivity. The hypothesis is proposed, Riemann metric is only possible for
quantized space classes. This statement probably can close the problem of metrical
form. 
\end{abstract}
During long time the space problem is discussed in many works from the point
of view of connection between space and time coordinates\cite{c1}. It is reckoned,
the metrical space form is proved by experimental data and this question was
not discussed. But now many virtual spaces exist ( see, for example \cite{c2}
) and there is not guaranty, that space metrical form is unique. 

The following approach probably closes the metrical form problem. Certainly,
all spectrum of this problem is not discussed in this article , particularly
, in cases, when global constants can be changed, etc, but the proposed method
can be useful in future.

As known, Fermat's theorem is formulated\footnote{
See information about  proof in \cite{b1} .
} as the follow statement: \textit{For any integer nonzero N\( _{x} \), N\( _{y} \),
N\( _{L} \)}\footnote{
In the most part, it's written\textit{''for natural number '',} but for ``
\textit{nonzero integer number}'' theorem is correct too.
} \textit{and natural \( \alpha \geq 3 \), it is not possible the follow equality:}

\begin{equation}
\label{a1}
N^{\alpha }_{x}+N^{\alpha }_{y}=N^{\alpha }_{L}
\end{equation}

Using Plank's hypothesis \cite{b2}about space quantum\footnote{
The space quantum is \( q=\sqrt{\frac{Gh}{c^{3}}} \)=1.616\( \cdot 10^{-33} \)cm,
where \textit{G} is gravity constant, \textit{h} is Planck's constant, \textit{c}
is light beam speed.
}: \textit{The minimal value of space length exists:}

\begin{equation}
\label{a2}
L=Nl_{Pl}
\end{equation}

where \textit{N} is \textit{}natural, \textit{l\( _{Pl} \)} is space quantum. 

For the spaces having natural metrical degree the follow theorem can be proved:

\textbf{\textit{Theorem.}} \textit{If ( \ref{a1} ) and ( \ref{a2} ) are correct,
then the space metrical degree of quantized space have value \( \alpha \leq 2 \):}

\begin{equation}
\label{a3}
\left\Vert \mathbf{L}\right\Vert ^{\alpha }=\sum _{i}\left\Vert \mathbf{x}_{i}\right\Vert ^{\alpha }
\end{equation}

The proof is made by the substitution \( \left\Vert \mathbf{x}_{i}\right\Vert =N_{x_{i}}l_{Pl} \)
to ( \ref{a3} ), then equation ( \ref{a3} ) reduces to ( \ref{a1} ).

For \textit{L\( ^{p} \)} Banach's space the theorem is proved.

Unfortunately, this approach has one shortcut: the metrics is limited by the
form ( \ref{a3} ) only.

Let the vector \textbf{\textit{L}}, which be decomposed as

\begin{equation}
\label{a9}
\mathbf{L}=x\mathbf{i}+y\mathbf{j}
\end{equation}

and the metric be some function of \( \left\Vert \mathbf{x}\right\Vert  \),\( \left\Vert \mathbf{y}\right\Vert  \):
\( M=F(\left\Vert \mathbf{x}\right\Vert ,\left\Vert \mathbf{y}\right\Vert ) \)
.

In the begin, we make the assumption, which should be definitelty natural for
the physical space: the ``coordinate spliting''.

\textbf{\textit{Axiom.}} \textit{A coordinate system exists, where metric is
additive:}

\begin{equation}
\label{a4}
F(\left\Vert \mathbf{x}+\mathbf{y}\right\Vert =F(\left\Vert \mathbf{x}\right\Vert )+F(\left\Vert \mathbf{y}\right\Vert )
\end{equation}

or

\begin{equation}
\label{a5}
F(\left\Vert \mathbf{L}\right\Vert )=F(\left\Vert \mathbf{x}\right\Vert )+F(\left\Vert \mathbf{y}\right\Vert )
\end{equation}

Assuming Tailor's decomposition is possible for metric

\begin{equation}
\label{a6}
F(\left\Vert \mathbf{L}\right\Vert )=\sum _{i}a_{i}\left\Vert \mathbf{L}\right\Vert ^{i}=\sum a_{i}N_{i}^{i}l^{i}_{Pl},
\end{equation}

and using Borh's statement about the equivalence of continuous and discrete
approaches and accounting ( \ref{a6} ), we can rewrite ( \ref{a5} ) as:

\[
a_{1}N_{L}=a_{1}N_{x}+a_{1}N_{y}\]

\[
a_{2}N^{2}_{L}=a_{2}N^{2}_{x}+a_{2}N^{2}_{y}\]

\begin{equation}
\label{a7}
..................................
\end{equation}

\[
a_{i}N^{i}_{L}=a_{i}N^{i}_{x}+a_{i}N^{i}_{y}\]

\[
..................................\]

( the grouping is made for degrees of \textit{l\( _{Pl} \)} , taking into account
that \( \left\Vert \mathbf{x}\right\Vert  \), \( \left\Vert \mathbf{y}\right\Vert  \)
are arbitrary ).

But \ref{a7} is possible only if \( i\leq 2 \) ( see previous theorem ) or
for \textit{a\( _{i} \) =} 0.

This means that:

\begin{equation}
\label{a8}
M(\left\Vert \mathbf{L}\right\Vert )=a\left\Vert \mathbf{x}\right\Vert ^{2}+b\left\Vert \mathbf{x}\right\Vert 
\end{equation}

For our case \textit{b =} 0 . 

So we nearly proved

\textbf{\textit{Hypothesis..}} \textit{Any additive metric in quantized space
is quadratic.}

\textbf{\textit{Remark 1}}

It is interesting, that \textit{a} can be zero. One-dimensional space does not
have decomposition ( \ref{a9} ). However, for any \textbf{a} , \textbf{b} in
this space \textbf{}and \textbf{c}=\textbf{a}+\textbf{b} expression ( \ref{a5}
) is valid too: ( \ref{a5} ) does not rigoursly demand the decomposion ( \ref{a9}
), ( \ref{a4} ) is enough. 

\textbf{\textit{Remark 2.}}

In the framework of Borh's conception \cite{b3} we used ( \ref{a1} ), ( \ref{a2}
) and ( \ref{a7} ), that is not quite consistent , if one follows formal mathematical
logic ( \textit{N \( _{x} \), N}\( _{y} \) are not generally arbitrary for
discrete approach). This contradiction is escaped, because the numbers \textit{N
\( _{x} \), N}\( _{y} \) take ''almost'' all values: almost for any great
value \textit{N}\( _{L} \) we can find \textit{N \( _{x} \), N}\( _{y} \)
which satisfy to ( \ref{a1} ). Their number is infinite, furthermore, they
have the very dense spectrum of values, that is enough for the space construction.
In this case, ( \ref{a8} ) can be used.

\textbf{\textit{Remark 3.}}

The proposition, that linear operator exists, was not used: we did not inroduce
scalar multiplicity definition, only special system, where metrics is additive.
This indicates , our space needs not be Hilbert's. 

\textbf{\textit{Conclusion.}}

For the quantized space it is enough to assume that the space system, having
the additive metric, exists. Then we can conclude, that our space is Hilbert's.

So for the space coordinates the connection between SQE and the metric form
is proved. 

For the space, which includes the time, the problem is known can be solved on
the basis, that space coordinates' metric is quadratic \cite{c1} . This definitely
means that pseudoeuclidic metric for inertial coordinate systems and, consequently,
Riemann's metric for non-inertial system\cite{c1} is only possible for quantized
spaces classes. 

These theorems are also the additional indirect proof of SQE.

Author would like thank I.Koganov for the discussion.


\begin{thebibliography}{}
\bibitem{c1}L.D. Landau, E.M. Lifshitz. The classical theory of fields, Oxford, Pergamon
Press, 1975. 
\bibitem{c2}R. E. Megginson. An introduction to Banach space theory, New York, Springer,
1998.
\bibitem{b1}S. Singh. Fermat's enigma: the epic quest to solve the world's greatest mathematical
problem, New York, Walker, 1997.
\bibitem{b2}M.K.E.L. Planck. Deut. Akad. Wiss., Berlin, K1, Math.-Phys. Tech., 440 ( 1899
). 
\bibitem{b3}A. Whitaker. Einstein, Bohr and the quantum dilemma, Cambridge, Cambridge University
Press,1996.
\end{thebibliography}
\end{document}